\documentstyle[twocolumn]{jpsj}
\def\runtitle{Interaction effects in multi-subband quantum wires}
\def\runauthor{Arisato Kawabata, Tobias Brandes}
\title{Interaction effects in multi-subband quantum wires}
\author{Arisato Kawabata, Tobias Brandes}
\inst{Department of Physics, Gakushuin University
1-5-1 Mejiro, Toshima-ku, Tokyo 171 Japan}
\recdate
{
\today
}
\abst{ 
We investigate the effect of electron-electron interactions on the transport 
properties of disordered quasi one-dimensional quantum wires with two or more 
subbands occupied. We apply two alternative methods to solve the logarithmic 
divergent problem, namely the parquet graph theory and a renormalization 
group calculation. We solve the group equations analytically in the weak 
coupling limit and find a power-law for the temperature dependent 
conductivity of a multi-channel system. The exponent is roughly equal to the 
inverse of the number of the occupied subbands. 
} 
\kword
{
Tomonaga--Luttinger liquid, quantum wire, Coulomb interaction, mesoscopic, 
renormalization group
}

\begin{document}
\maketitle

One-dimensional interacting electron systems have attracted considerable 
interest recently. Much attention has focussed on  
striking effects in presence of disorder potentials such as a vanishing 
conductance at zero temperature and non-Luttinger liquid behavior 
\cite{kaneandglazman}. Experimentally, the temperature dependence of the 
conductivity $\sigma (T)$ of long wires \cite{THS95} seems to favor 
predictions of the Tomonaga-Luttinger model \cite{Voi95}. Although the 
latter has become a standard way to describe correlated 1d electrons, there 
are still arguments for ordinary Fermi liquid theory to be valid in the 
presence of disorder \cite{HDS93}. Investigations of multi-component models 
where directed mainly towards a small number of coupled chains \cite{Voi95}, 
triggered by the possible relation to 2d superconductivity \cite{And91}, or 
tunneling \cite{MG93}. On the other hand, in transport experiments one is 
able to drive a system from the extreme quantum limit of one to a large 
number of $N$ occupied subbands. The question then is how Coulomb effects 
at low temperatures change when $N$ is varied, with the possible re-
establishing of Fermi-liquid behavior in the limit of large $N$. 

In this Letter, we present a microscopic theory for the conductivity $\sigma$ 
of a multi- subband wire in presence of an arbitrary, weak static disorder 
potential and Coulomb interactions between the electrons. One of our central 
predictions is a temperature dependence 
\begin{equation}
\label{prediction}
\sigma (T)\sim T^{1/N}
\end{equation}
in the limit of weak interactions $a^{*}/W\gg 1$, where $a^{*}$ is the 
effective Bohr radius and $W$ the wire width.                       
A similar $1/N$ dependence has been found by  Flensberg \cite{Fle93} for the 
conductance of a quantum dot coupled to quantum point contacts in the 
framework of a multichannel Tomonaga-Luttinger model. 
The result Eq.~(\ref{prediction}) is a generalization of the one-cannel case 
\cite{Kaw94} $ \sigma (T)\sim T^{\lambda (2g_{2}-g_{1})}$, where the factor 
$\lambda g_{2}\approx 1/2$ turned out to be 'nearly' universal, while 
$\lambda g_{1}=c_{1}/a^{*}k_{F}$ ($c_{1}$ a constant of order unity which 
depends on the structure of the wire) is negligeable since typically 
$k_{F}\sim 1/W$. 

We generalize the parquet diagram theory \cite{BGD66,Kaw94} to the present 
case and give an alternative, microscopic derivation using a standard one-
loop renormalization group (RG) calculation which reproduces exactly the 
vertex equations of the parquet theory. It turns out that the singular ($q\to 
0$), logarithmically divergent Coulomb interaction has to be renormalized. 
Namely, one has to take into account all diagrams which contain powers of the 
interaction matrix elements, which physically means that the interaction 
effectively is replaced by a screened Coulomb interaction. This reduces the 
problem to that of short range interactions which dominate the low-$T$ 
behavior of the conductivity. The latter conclusion is in contrast to 
calculations \cite{SchulzMaurey}, where a contribution from $4k_{F}$ 
scattering was found to be dominant at low $T$ in the case of long range 
Coulomb interactions and delta impurities. 
At present, it is not clear if 
this 'Wigner-crystal pinning' effect dominates over the $2k_{F}$--
contribution in experiments with relatively short wires \cite{THS95}. 
Nevertheless, we believe that there should be a wide temperature regime over 
which the $1/N$ behavior Eq.~(\ref{prediction}) can be observed. 

We start from a multi-band quantum wire with dispersion $E_{nk}$,
where $n$ is the band index and $k$ the momentum of plane waves 
in $x$-direction. 
The interaction potentials are $g_{nm}(q)$ for scattering of 
two incoming electrons in band $n$ to band $m$,
and $h_{mn}(q)$ for scattering of two electrons from
band $m$ and $n$ to band $n$ and $m$ with momentum transfer $q$,
\begin{subeqnarray}
\label{interaction}
\frac{\varepsilon^*{g}_{nm}(q)}{2e^{2}}&\!\!\!\!\!\!\!=&\!\!\!\!\!\!\!\!\!
\int\!dydy'\!
{K}_{0}(q|y\!-\!y'|){\phi }_{m}(y){\phi }_{n}(y'){\phi }_{m}(y'){\phi
}_{n}(y),\nonumber\\
\\
\frac{\varepsilon^*{h}_{nm}(q)}{2e^{2}}&\!\!\!\!\!\!\!=&\!\!\!\!\!\!\!\int
dydy'\!
{K}_{0}(q|y\!-\!y'|){\phi }_{m}^{2}(y){\phi }_{n}^{2}(y'),
\end{subeqnarray}
where $K_{0}(z)$ is the modified Bessel function, $\varepsilon ^{*}$ the 
effective dielectric constant, and $\phi_{n}(y)$ the (real) wave functions in 
transversal direction.
The vertex  corrections to 
the backward scattering are singular when the wave numbers of the incoming and 
outgoing electrons are close to one of the Fermi wave numbers $\pm k_{n}$,
while those
to the 
forward scattering are singular only through the divergence of 
the potentials at small $q$ \cite{BGD66,Kaw94}. For the moment, we assume
short range 
potentials $g_{nm}$ and $h_{nm}$ regular at $q=0$ and show below how to 
treat the Coulomb case.
We write the vertices in the form
${\Phi }_{s}({{\mib p}}_{1},{{\mib p}}_{2},{{\mib p}}_{3},{{\mib p}}_{4})$,
where ${\mib p} = (p_{i}, \varepsilon_{i} )$,
$p_{1}$, $\varepsilon _{1}$ and $p_{2}$, $\varepsilon _{2}$ are 
the wave numbers and the
Matsubara frequencies 
of the incoming electrons,
$p_{3}$, $\varepsilon _{3}$ and $p_{4}$, $\varepsilon _{4}$ are those of the
outgoing
electrons, and the 
suffix $s$ indicates the spins states of the incoming electrons, i.e., $s = 
1$ for anti-parallel 
spins and $s =2$ for parallel spins.  
Two kinds of vertices have to be distinguished. The vertex 
$\gamma _{snm}({\mib p}_{1}, {\mib p}_{2},{\mib p}_{3}, {\mib p}_{4})$ 
$(n\ne m)$ is that for $p_{1}\cong  -k_{m}$, $p_{2}\cong  k_{n}$,
$p_{3}\cong  k_{n}$, $p_{4}\cong -k_{m}$, and $\chi_{snm}(
{\mib p}_{1}, {\mib p}_{2},{\mib p}_{3}, {\mib p}_{4})$ is that 
for $p_{1}\cong -k_{n}$, $p_{2}\cong  k_{n}$,
$p_{3}\cong  k_{m}$, $p_{4}\cong -k_{m}$
(the case of $n = m$  will be included in 
$\chi_{snm}({\mib p}_{1}, {\mib p}_{2},{\mib p}_{3}, {\mib p}_{4})$). 
In the parquet diagram 
method \cite{BGD66}, the 
vertex $\gamma _{snm}$ is composed of two irreducible parts 
$\sigma_{snm}$ and $S_{snm}$: $\sigma_{snm}$ ($S_{snm}$) 
is the sum of the all the graphs which can be separated by cutting two 
antiparallel (parallel)  electron lines. In the same way  
irreducible 
vertex parts $\tau_{snm}$ and $T_{snm}$ for $\chi_{snm}$ are introduced. 
The corresponding integral equations have the form $(m\ne n)$ 
\begin{eqnarray}
& &{\sigma }_{1nm}({{\mib p}}_{1},{{\mib p}}_{2},{{\mib 
p}}_{3},{{\mib p}}_{4})= {g}_{1nm}\nonumber\\
&+& {k}_{B}T\sum_{\mib 
k} \{{S}_{1nm}({{\mib p}}_{1},{\mib k},{{\mib p}}_{3},{\mib k}-{\mib 
q}) {G}_{n}({\mib k}){G}_{m}({\mib k}-{\mib q})
\nonumber\\
&\times&\left.{\gamma }_{2nm}({\mib 
k}-{\mib q},{{\mib p}}_{2},{\mib k},{{\mib p}}_{4}) +   
{T}_{1mn}({{\mib p}}_{1},{\mib k},{{\mib p}}_{3},{\mib k}-{\mib q})\right.
\nonumber\\
&\times&{{G}_{m}({\mib k}){G}_{n}({\mib k}-{\mib q}){\chi }_{2nm}({\mib k}-
{\mib q},{{\mib p}}_{2},{\mib k},{{\mib p}}_{4})+ 
(1\leftrightarrow 2)}
\}, 
\nonumber\\
\label{(3.2)} 
\end{eqnarray}
where ${\mib q} ={\mib p}_{3} -{\mib p}_{1}={\mib 
p}_{2}-{{\mib p}}_{4}$, ${g}_{1nm}= {g}_{nm}({k}_{n}+{k}_{m})$, and 
${G}_{n}({\mib k})$ is the one electron Green's function 
$G_{n}({\mib k})= 1/(i\hbar\varepsilon _{\nu }-{E}_{n,k}+ \mu)$, 
$\mu$ being the chemical potential. The notation 
$(1\leftrightarrow 2)$ means that the terms in which the subscript $1$ and 
$2$ of the vertices are 
interchanged are to be added. 
The sum $\sum_{{\mib k}}=\sum_{\varepsilon _{\nu }}\int dk/2\pi$ 
gives rise to the logarithmic divergence at zero temperature when ${\mib q} =
 (k_{n} + k_{m} , 0)$.  
The parquet graph method takes into account the contributions of all graphs 
giving rise to such singularities. The 
vertices can be regarded as functions of two parameters
$\eta$ and $\xi $, where ($v_{n}$ are the Fermi velocities)
\begin{equation}
\label{eta}
\eta \equiv {\frac{1}{4}}\log\left[{{\frac{{\varepsilon
}_{0}^{4}}{({\omega }^{2}+{v}_{n}^{2}\Delta {q}^{2})({\omega
}^{2}+{v}_{m}^{2}\Delta {q}^{2})}}}\right] ,
\label{(3.6)}
\end{equation}
and a corresponding definition for $\xi $ with $(\Delta q,
\omega )$ replaced by $(\Delta p,\Delta \varepsilon )$, where $\Delta q=
{p}_{3} - {p}_{1} - {k}_{m} - {k}_{n}$, $\omega =
{\varepsilon }_{3} - {\varepsilon }_{1}$, $\Delta p= {p}_{1}+ {p}_{2}+
{k}_{m} - {k}_{n}$ and $\Delta \varepsilon = {\varepsilon }_{1}+
{\varepsilon }_{2}$, 
$\varepsilon _{0}$ being the cut-off frequency of order of 
Fermi energy devided by $\hbar$.
Thus one has ${\sigma }_{1nm}({{\mib p}}_{1},{{\mib p}}_{2},{{\mib
p}}_{3},{{\mib p}}_{4})
=
{\sigma }_{1nm}(\xi ,\eta )$ 
etc. We introduce 
${\lambda }_{nm}:= (\pi\hbar({v}_{n}+{v}_{m}))^{-1}$
and the product
$F\cdot G(\xi ,\eta )= \int_{0}^{\xi }F(\xi ,
\min[t,\eta])G(\xi ,t)dt$
for $\xi\!>\!\eta$ and $F\cdot G(\xi ,\eta )=\int_{0}^{\eta}F(\min[t,\xi ]
,\eta )G(t,\eta )dt$ for $\eta\!>\!\xi$.
In the case $\eta\!>\xi$, eq.~(\ref{(3.2)}) can be written  
in the form
\begin{eqnarray}
{\sigma }_{1nm}(\xi ,\eta )&-&{g}_{1nm}=-{\lambda
}_{nm}\left\{{{S}_{1nm}\cdot {\gamma }_{2nm}(\xi ,\eta )+ }\right.\nonumber\\
& &\left.+{T}_{1mn}\cdot {\chi }_{2nm}(\xi ,\eta )+ (1\leftrightarrow 2)
\right\}.
\label{(3.10)}
\end{eqnarray}
The equations for the remaining vertices are similar except for 
the spin structures and the terms of lowest order in the interaction. Since 
the expressions are quite lengthy, we will present the detailed calculations 
elsewhere. The important point is 
that the set of integral equations can be transformed into differential 
equations whose physical meaning becomes transparent in view of the 
RG calculation below. The result is
\begin{subeqnarray}
\label{lieequations}
{\gamma }_{1nm}'(\xi )&\!\!\!\!\!\!\!=&\!\!\!\!\!\!\! -2{\lambda
}_{nm}\left[{{\gamma }_{1nm}^{2}(\xi )+ {\chi
}_{1nm}(\xi ){\chi }_{2nm}(\xi )}\right],\\
{\gamma }_{2nm}'(\xi )&\!\!\!\!\!\!\!=&\!\!\!\!\!\!\! -{\lambda
}_{nm}\left[{{\gamma }_{1nm}^{2}(\xi )+
{\chi }_{1nm}^{2}(\xi )+ {\chi }_{2nm}^{2}
(\xi )}\right], \\
{\chi }'_{1nm}(\xi )&\!\!\!\!\!\!\!=&\!\!\!\!\!\!\! -2{\lambda
}_{nm}\left[{{\gamma
}_{1nm}(\xi ){\chi }_{2nm}(\xi )
+{\gamma }_{2nm}(\xi ){\chi }_{1nm}(\xi )}\right]\nonumber\\
+\sum\limits_{l}^{} &\!\!\!\!\!\!\!{\lambda
}_{ll}&\!\!\!\!\!\!\!\left[{{\chi }_{1nl}(\xi ){\chi
}_{2lm}(\xi )+ (1\!\leftrightarrow\!2)
-2{\chi }_{1nl}(\xi ){\chi }_{1lm}(\xi )}\right],\nonumber\\
\\
{\chi }_{2nm}'(\xi )&\!\!\!\!\!\!\!=&\!\!\!\!\!\!\! -2{\lambda }_{nm}
\sum_{s=1,2}{\gamma}_{snm}(\xi ){\chi }_{snm}(\xi )\nonumber\\
&\!\!\!\!\!\!\!+&\!\!\!\!\!\!\! \sum\limits_{l}^{} {\lambda }_{ll}{\chi
}_{2nl}(\xi ){\chi
}_{2lm}(\xi ),\quad (n\ne m) \\
{\chi }_{1nn}'(\xi )&\!\!\!\!\!\!\!=&\!\!\!\!\!\!\! -2\sum\limits_{l}^{} {\lambda
}_{ll}{\chi }_{1nl}^{2}(\xi )+ 2\sum\limits_{l\ne
n}^{} {\lambda }_{ll}{\chi }_{1nl}(\xi ){\chi }_{2ln}(\xi ),\nonumber\\
\\
{\chi }_{2nn}'(\xi )&\!\!\!\!\!\!\!=&\!\!\!\!\!\!\! -{\lambda }_{nn}{\chi
}_{1nn}^{2}(\xi )+ \sum\limits_{l\ne n}^{} {\lambda
}_{ll}{\chi }_{2nl}^{2}(\xi ).
\end{subeqnarray}
The initial values of the vertex functions are
${\gamma }_{1nm}(0)= \chi _{1nm}(0)=g_{1nm}$, 
${\gamma }_{2nm}(0)= g_{1nm}-h_{2nm}$, and ${\chi}_{2nm}(0)= g_{1nm}-
g_{2nm}$.

We now outline the RG calculation.
There, it turns out that the case $\eta = \xi$, eq.~(\ref{eta}),
namely $\Delta q=\Delta p=0$, and $\Delta \varepsilon =\omega =\varepsilon 
_{1}-\varepsilon _{3}=\varepsilon _{1}+\varepsilon _{2}$, is sufficient
to extract the logarithmic divergence of the vertex and to reproduce
eqs.~(\ref{lieequations}).
If the energy variables are chosen as in \cite{Sol79}, $\varepsilon  
_{1}=(3/2)\omega $, $-\varepsilon  _{2}=\varepsilon  _{3}= \varepsilon  _{4}= 
(1/2)\omega $, the vertex depends on one single variable $\omega $ only (the 
energy transfer in a scattering event) which, however, is sufficient to 
extract the low-temperature dependence from a $T=0$ calculation via 
$\hbar\omega \to k_{B}T$. 
The diagrams contributing to $\gamma_{snm}$ and $\chi _{snm}$ to 
lowest order are the standard bubble, triangle, ladder and 
crossed types. For their evaluation,
we define Cooper and zero sound functions
\begin{eqnarray}
C_{mn}(\omega ,q):=\int\frac{dkd\omega '}{(2\pi)^{2}}
G_{m}(k,\omega ') G_{n}(q-k,\omega -\omega ') 
u(k),\nonumber\\
\end{eqnarray}
and correspondingly $Z_{nm}(\omega ,q)$ with $(q-k, \omega -
\omega ')$ replaced by $(k+q,\omega '+\omega )$. Here,  
$u(k)$ is an arbitrary non-singular function, and 
$G_{n}(k,\omega )$ is the free $T\!=\!0$-Green's function. 
One can easily extract the logarithmic divergent part 
for small momentum transfers, $C_{mn}(\omega ,q=k_{n}-k_{m})\cong -i
\left[u(-k_{m})+\delta _{nm} u(k_{n})\right]
\lambda_{nm}\ln (\omega/\varepsilon_{0})$.
For $q=k_{n}+k_{m}$, one has
$Z_{nm}(\omega ,q)\cong iu(-k_{n})\lambda_{nm}\ln (\omega/\varepsilon_{0})$.
In both cases, the logarithmic divergence is due to a small region in $k$-
space around the points $\pm k_{n},\pm k_{m}$, where the dispersion 
$E _{nk}$ can be linearized. In the Tomonaga-Luttinger 
model this linearization is performed from the very beginning, here it 
appears naturally in the evaluation of the diagrams. 
Careful evaluation yields ($n \ne m$)
\begin{eqnarray}
\label{chinm}
\chi_{1nm}(\omega )=g_{1nm}
-\ln \frac{\omega }{\varepsilon_{0}}\left\{\phantom{\sum_{l}
\!\!\!\!\!\!\!}
2\lambda _{mn}[
g_{1nm} h_{2nm}+(1\leftrightarrow 2)]
\right.\nonumber\\
-\left.4\lambda _{mn}g_{1nm} h_{1mn}
-\sum_{l}\lambda _{ll}\left[g_{2nl} g_{1lm} +g_{1nl} g_{2lm}\right]
\right\},{\hspace{0.5cm}}
\end{eqnarray}
and similar expressions for the other vertices. 
Here, for simplicity we did not include a spin-dependence of the interaction
parameters or the $\lambda _{mn}$ as, e.g., in a spin-orbit split system or 
in presence of a magnetic field, although the formalism 
easily incorporates these effects.  

In the renormalization procedure,
the cutoff $\varepsilon_{0}$ provides a natural means to scale to the new set 
of models \cite{Sol79} defined by a smaller value 
$\varepsilon_{0}'<\varepsilon_{0}$. At the same time, 
one requires invariance of the scattering properties of the system which means 
invariance of the vertices, e.g. in the case of $\chi _{1nm}$ 
\begin{equation}
\label{renorm1}
\chi_{1nm}(\omega,\Gamma _{nm}(\varepsilon_{0}))=\chi_{1nm}\left
(\frac{\varepsilon_{0}}{\varepsilon_{0}'}\omega,\Gamma
_{nm}(\varepsilon_{0}')\right),
\end{equation}
where $\Gamma _{nm}(\varepsilon_{0})$ denotes the coupling matrix 
with the corresponding 'initial' conditions $g_{snm}(\omega\!=\!\varepsilon_{0})
\!=\!g_{snm}$ 
and $h_{snm}(\omega\!=\!\varepsilon_{0})\!=\!h_{snm}$. 
The functional dependence of the vertex on the energy and 
the coupling constants thus remains invariant under a scale transformation. 
For higher than second order perturbation theory in the bare 
couplings, one should consider the product of the vertex with two of its 
'arms' (single-particle Green's-functions) as invariant, rather than the 
vertex itself \cite{Sol79}. However, since self--energy corrections $\Sigma $ 
are divergent only in second order and the bare 
vertex itself already is first order, eq.~(\ref{renorm1}) is consistent. In 
fact, the first order Hartree-Fock self--energy $\Sigma _{n}(k)$ 
merely renormalizes the band energies $\xi _{nk}$ (counted from the chemical 
potential), namely 
$\xi_{nk}= \varepsilon _{nk}+\Sigma _{n}(k)-\mu$, 
with the Fermi momenta $k_{n}$ as solutions of $\xi _{nk_{n}}=0$.

The dependence of the coupling parameters on $\varepsilon_{0}'$ gives the 
flow to the new set of models with modified cutoff. 
By differentiating eq.~(\ref{renorm1}) with respect to 
$\omega $ and setting the new cutoff $\varepsilon_{0}'=\omega $,
\begin{equation}
\chi'_{1nm}(\omega)=(\varepsilon_{0}/\omega)\frac{\partial}{\partial \Omega }
\left.\chi_{1nm}\left
(\Omega ,g(\omega )\right)\right|_{\Omega =\varepsilon_{0}},
\end{equation}
one obtains one of the RG equations eqs.~(\ref{lieequations}), and 
correspondingly for $\gamma _{2nm}(\omega )$ etc. Frequency dependent 
singlet- and triplet vertices correspond to the respective couplings in the 
RG equations, $\gamma_{1mn}(\omega )=h_{1mn}(\omega )$,$ \gamma _{2mn}(\omega 
)=h_{1mn}(\omega ) -h_{2mn}(\omega )$,$\chi _{1mn}(\omega )=g_{1mn}(\omega 
)$, and $\chi_{2mn}(\omega )=g _{1mn}(\omega )-g_{2mn}(\omega )$. One can 
easily verify the agreement with eqs.~(\ref{lieequations}) upon introducing 
the dimensionless logarithmic variable $\xi =-\ln (\omega /\varepsilon_{0})$. 

We now turn to the conductivity $\sigma $ of the multi-subband wire, namely its 
temperature dependence, where the power-law due to the interaction 
effects is predicted. The latter are expected to be important in relatively 
clean wires, where the impurity potential $V_{nm}(q)$ for scattering from 
band $n$ to $m$ with momentum transfer $q$ can be treated as a 
perturbation. However, the screening by the interacting
electron gas is non-perturbative and changes the 
impurity (backscattering) $V_{nm}(k_{n}+k_{m},T)\equiv V_{nm}(T)$,
which becomes temperature dependent and divergent 
for $T\to 0$, drastically.
In the parquet method, upon inserting the full backward scattering 
vertices, an integral equation for $V_{nm}(T)$ is obtained describing the 
$T$-dependent screening effects on the impurity potential.
In the equivalent RG calculation, one first starts from 
the perturbative expression
for $V_{nm}(\omega )$ which again is logarithmic divergent, so that a scaling 
procedure must be performed, namely 
\begin{equation}
V_{nm}(\omega,\Gamma_{nm} 
(\varepsilon_{0}))=z(\frac{\varepsilon_{0}}{\varepsilon_{0}'}) 
V_{nm}\left((\frac{\varepsilon_{0}}{\varepsilon_{0}'})\omega, \Gamma_{nm} 
(\varepsilon_{0}')\right). 
\end{equation}
Here, scaled and unscaled quantities differ
by a multiplicative factor $z$ independent of $\omega $.
The bare couplings are replaced by the vertices to include all possible 
intermediate interaction-induced scattering processes which leads to         
\begin{subeqnarray}
\label{vdifferential}
\frac{\partial}{\partial \omega }\ln V_{nm}(\omega )=-\frac{\alpha
_{nm}(\omega )}{\omega },
\hspace{2cm}\\ 
V_{nm}(\varepsilon_{0})=V_{nm},\hspace{3cm}\\
\alpha _{nm}(\omega ):=-\lambda _{nm}\sum_{s=1,2}\left[\gamma _{smn}(\omega )+
\chi _{snm}(\omega )\right],
\end{subeqnarray}
for $n\!\ne\!m$, where we defined $V_{nm}:=V(k_{n}+k_{m})$, and
$\gamma_{nm} \equiv 0$ for the case $n\!=\!m$.
For low temperatures or correspondingly small $\omega $, one can 
replace the vertices by their fixpoints, i.e their
$\xi\! \to\! \infty (\omega \to 0)$ values.
The solution gives the temperature dependence upon replacement of $\omega $ 
by $k_{B}T/\hbar$,
\begin{equation}
\label{VTapprox}
V_{nm}(T)=V_{nm}\cdot(T/T_{0})^{-\alpha _{nm}}, 
\end{equation}
where $\!\alpha _{nm}:=\alpha_{nm}(\omega=0)$ and $T_{0}:=\varepsilon_{0}/k_B$.
Here, we  neglected logarithmic corrections of the form $\log( T/T_{0})$ which 
experimentally  are not very important compared to the power-law dependence 
of the conductivity $\sigma $ to be derived now. 

Since the impurity potential is treated perturbatively, $\sigma $ can be
obtained from the Boltzmann equation
\cite{AA91} for the multi-channel case, 
\begin{subeqnarray} \sigma 
= {\frac{2{e}^{2}}{\pi \hbar}}\sum\limits_{n=1}^{N} {l}_{n} ,\hspace{1.5cm}\\
\sum\limits_{m=1}^{N} ({K}_{nm}-{J}_{nm}){l}_{m}= 1, \label{(6.3)} 
\end{subeqnarray} 
where $L K_{nm}=m^{*2}|{V}_{nm}(T){|}^{2}/(k_{n}k_{m}\hbar^{4})$, $L 
J_{nm}=m^{*2}\times $
$|{V}_{nm}(k_{n}-k_{m}){|}^{2}/(k_{n}k_{m}\hbar^{4})$ ($m\ne n$), and 
${J}_{nn}=$ 
$-\sum_{m\ne n}$ ${J}_{nm}-\sum_{m}{K}_{nm}$, $L$ 
being the length of the wire, and $m^{*}$ the electron band mass. 
In the above, we used the bare scattering amplitudes for the intersubband
forward 
scatterings since their temperature dependences are not strong.
Below we assume that $\alpha _{nm}> 0$ for all $n$ and $m$. In fact it is
the case for
all the 
examples of our numerical calculations discussed below. Then, at
very low 
temperatures the $K_{nm}$ are divergent like ${K}_{nm}\propto{T}^{-2{\alpha
}_{nm}}$, and we may neglect $J_{nm}$   
except $J_{nn}$ which contains $K_{nm}$.  
We obtain
$\sum_{m=1}K_{nm}(l_m+l_n)= 1$.                                     
Note that each term in the left hand side of this equation is
positive and 
hence has to remain finite in the limit of zero temperature
although the $K_{nm}$ 
are divergent.
Let $M(n)$ be the value of $m$ for which $\alpha _{nm}$ is the largest for 
given $n$, and  $\alpha _{n} 
= \alpha _{nM(n)}$ . Then, assuming that 
$
{l}_{n}\propto {T}^{2{\alpha }_{n}}$ 
at very low $T$, the term 
$\sum_{m=1}{K}_{nm}l_{n}$ tends to a 
non-vanishing constant in the limit of zero temperature. As for the first
term $\sum_{m=1}{K}_{nm}l_{m}$, we 
find that $K_{nm} = K_{mn}$, and hence that $K_{nm}l_{m}$
remains finite 
because $K_{nm}$ diverges more slowly than or like ${T}^{-2{\alpha }_{m}}$. Thus
the assumption 
${l}_{n}\propto {T}^{2{\alpha }_{n}}$
is reasonable, and the conductivity
behaves like
$
\sigma \propto {T}^{2\alpha },
$
at very low temperatures, where $\alpha $ is the smallest of the 
$\alpha_{n}$.

So far we have not taken into account the divergence of the
bare interaction potential at small $q\to0$, see eqs.~(\ref{interaction}).
The elements $g_{nn}(q)$ and $h_{nm}(q)$ are divergent like    
${h}_{nm}(q)= {g}_{nn}(q)\approx-
(2{e}^{2}/\varepsilon^*)\log Wq\equiv {U}_{c}(q)$ for
$q \rightarrow  0$, where $W$ is of the order of the width of the wire.
This divergence is of the same kind as the one which
appeared in the self-consistent equations for the vertices. 
Therefore in diagrams containing $g_{nn}(q)$ and $h_{nm}(q)$ with 
small $q$ one has to replace the interaction with the screened Coulomb 
interaction ${U}_{sc}({\mib q})= {U}_{c}(q)/(1+{U}_{c}(q)\sum_{l=1}^{N}
{\mit\Pi}_{l}({\mib q}))$
where ${\mib q} = (q, \omega )$.
Here,
\begin{equation}
{\mit\Pi}_{l}({\mib q})= 2{v}_{l}{q}^{2}/[\pi\hbar({v}_{l}^{2}q^{2}+{\omega 
}^{2})]
,
\end{equation}
is the polarization function in subband $l$.
Typical  values of $\omega $ and $q$ 
for $g_{nn}(q)$ are those with $\omega \sim v_{n}q$. Since $q$
is very small, one can put
\begin{equation}
\label{(7.8)}
{g}_{nn}(0)^{-1}= \frac{2}{{\pi \hbar}}\sum_{l=1}^{N}
\frac{{v}_{l}}{{v}_{l}^{2}+{v}_{n}^{2}}.
\end{equation}
On the other hand, the situation for $h_{nm}(q)$ is not so simple; 
we find 
that the important value of $\omega$ is an average of
$v_{n}q$ and $v_{m}q$
which for $h_{nm}(0)$ leads an expression like eq.~(\ref{(7.8)})
with $v_{n}q\to(v_{n}\!+\!v_{m})q/2$. 
This is sufficient to study the dependence of
the exponent 
$\alpha _{nm}$ on $N$, the number of the occupied subbands,
since the problem reduces to that of short range 
interactions and we can apply the preceding results.

In order to obtain the exponents $\alpha_{nm}$, we have to solve the
coupled equations 
eqs.~(\ref{lieequations}). 
The fixpoints of the latter at $\xi \rightarrow  \infty $
are obtained by putting the 
right hand sides to zero. The initial values of 
$\gamma _{snm}$ and $\chi_{snm}$ are real, 
and hence they are real for arbitrary $\xi$. Then, from the equation for 
$\gamma _{2nm}(\xi )$
we find that
${\gamma }_{1nm}(\infty )= {\chi }_{1nm}(\infty )= {\chi }_{2nm}(\infty )=
0,(n\ne m )$ and hence ${\chi }_{1nn}(\infty )= 0$. 
Although the eqs.~(\ref{lieequations}) are  very complex, we can solve them in
some limit. Considering the expressions eqs.~(\ref{interaction}),
we first note that
unless $Wq \ll 1$, $h_{nm}(q)$ and $g_{nn}(q)$ are of order 
$e^{2}/\varepsilon ^{*}$. 
On the other hand, as
can be 
seen from eq.~(\ref{(7.8)}), $g_{nn}(0)$ and $h_{nm}(0)$ are of order of the
typical Fermi 
velocity times $\hbar$ . Therefore the ratios of $h_{nm}(0)$ or $g_{nn}(0)$ 
to $h_{nm}(q)$ or $g_{nn}(q)$ for $q\sim 
1/W$ are of order $a^{*}$ times the typical Fermi wave number, where 
$a^{*}\equiv 2\varepsilon^{*}\hbar/m^{*}e^{2}$  is the 
effective Bohr radius. The typical value of the Fermi wave number is of
order $1/W$, and 
those ratios are typically of order $a^{*}/W$, as are the ratios
of $\chi_{2nn}(0)$ and $\gamma_{2nm}(0)$ 
to the initial values of the other vertex functions. Thus when 
$a^{*}/W \gg 1$ (weak coupling limit), in the right hand side of 
eqs.~(\ref{lieequations}) 
we can neglect  terms containing neither $\chi _{2nn} (\xi)$ nor $\gamma
_{2nm}(\xi)$.
>From the remaining nonzero equations 
we find  $\chi _{snm}(\xi)$ to decrease exponentially if
${\lambda }_{nn}{\chi }_{2nn}(\xi )+{\lambda }_{mm}{\chi }_{2mm}(\xi
)-2{\lambda }_{nm}{\gamma }_{2nm}(\xi ) < 0$, 
where the weak coupling approximation is
consistent 
and valid. Then,  $\chi _{2nn}(\xi )$ and $\gamma 
_{2nm}(\xi)$
do not depend on
$\xi$, and 
with eq.~(\ref{VTapprox}) it follows ${\alpha }_{nm}\ \cong \ {\lambda 
}_{nm}{h}_{nm}(0)$ and ${\alpha }_{nn} \cong {\lambda }_{nn}{g}_{nn}(0)$. 
eq.~(\ref{(7.8)}) yields
${\alpha }_{nm}\ \cong \ {\alpha }_{nn}\ \cong \ 1/2N,
$
and with ${l}_{n}\propto {T}^{2{\alpha }_{n}}$ we arrive at our prediction 
eq.~(\ref{prediction}) $\sigma \sim T^{1/N}$. We note that the exponent $1/N$ 
no longer depends on the interaction parameters in the weak coupling limit.

In order to verify the validity of the latter,
we did numerical calculations for the case of two occupied subbands. 
The quantum wire was 
modeled by hard walls of width $W$. We calculated 
the quantities $\chi_{snm}(0)$ and
$\gamma _{snm}(0)$ 
and solved eqs.~(\ref{lieequations}) 
numerically.
Detailed results will be given elsewhere.
We used the values of 
$\gamma _{2nm}(\xi)$ and $\chi _{2nn}(\xi )$ for $\xi  = 10$ 
instead of the fixpoints $\gamma _{2nm}(\infty ) $
and $\chi _{2nn}(\infty )$. The $\chi _{snm} (\xi )$'s do not depend on $\xi $
very much up to $\xi  = 10$ and we find that 
the results are at large in agreement with 
${\alpha }_{nm}\ \cong \ {\alpha }_{nn}\ \cong \ 1/2N$.
Practically, the values of $\xi $ are limited by the temperature $T$ or by 
the length $L$  of the wires, $\xi  = 10$ corresponds to $10^{-4}$ 
times the typical Fermi temperature. If the latter is of order $100$ K, the
weak coupling approximation is valid down to $10$ mK.                      
         
Integrals over the wave number should be cut off at $1/L$ in the 
lower limit \cite{OF94}. This corresponds to a cutoff for 
$\xi $ at  ${\xi }_{L}\ \sim \ \log(1/{k}_{n}L)$, $k_{n}$ being the typical 
Fermi wave number. If $k_{n} = 10^{6}/m \sim 10^{5}/m$, $\xi _{L} = 10 $ 
corresponds to $L = 100\mu m\sim 10\mu m$ so that our 
results are applicable to real systems.

This work is supported by Grant-in-Aid for Scientific Research on Priority 
Areas, "Mesoscopic Electronics: Physics and Technology"
from the Ministry of Education, Science, Sports and Culture,
and a STF 9 European Union Fellowship.


\end{document}